\UseRawInputEncoding
\documentclass[showpacs,twocolumn,superscriptaddress,nofootinbib]{revtex4} 

\usepackage{doi}
\usepackage{hyperref}
\hypersetup{colorlinks=true,linkcolor=blue,citecolor=cyan}
\usepackage{graphicx}
\usepackage{xcolor, soul}
\sethlcolor{green}
\usepackage{dcolumn}
\usepackage{bm}
\usepackage{color}
\usepackage{enumitem}
\usepackage{mathrsfs}
\usepackage{amsmath}
\usepackage{amssymb}
\usepackage{cleveref}
\usepackage{orcidlink}

\crefname{equation}{Eqn.}{Eqns.}
\crefname{figure}{Fig.}{Figs.}
\crefname{section}{Sec.}{Sec.}
\crefname{table}{Table}{Tables}
\usepackage{physics}

\setstcolor{red}
 \definecolor{maroon}{rgb}{0.5, 0.0, 0.0}

\def\aap{Astronomy and Astrophysics}

\def\apjl{Astrophysical Journal Letters}

\def\apjs{The Astrophysical Journal Supplement}

\def\aap{Astron. Astrophys.}

\def\apjl{Astrophysical Journal Letters}

\def\apjs{The Astrophysical Journal Supplement}

\begin{document}

\title{%Energy 
Energetics of Buchdahl stars and the magnetic Penrose process}

\author{Sanjar Shaymatov
%\orcidlink{0000-0002-5229-7657}
}
\email{sanjar@astrin.uz}

\affiliation{Institute for Theoretical Physics and Cosmology, Zhejiang University of Technology, Hangzhou 310023, China}
\affiliation{Institute of Fundamental and Applied Research, National Research University TIIAME, Kori Niyoziy 39, Tashkent 100000, Uzbekistan}
\affiliation{Tashkent University of Applied Sciences, Gavhar Str. 1, Tashkent 100149, Uzbekistan}
%\affiliation{Western Caspian University, Baku AZ1001, Azerbaijan}

\author{Naresh Dadhich}
\email{nkd@iucaa.in}
\affiliation{Inter-University Centre for Astronomy \&
Astrophysics, (IUCAA), Post Bag 4, Pune 411 007, India}
\affiliation{Astrophysics Research Centre, School of Mathematics, Statistics and 
Computer Science, University of KwaZulu-Natal, Durban, South Africa }

\author{Arman Tursunov }
\email{atursunov@mpifr-bonn.mpg.de}
\affiliation{Max Planck Insititute for Radio Astronomy, Auf dem H{\"u}gel 69, Bonn D-53121, Germany}
\affiliation{Research Centre for Theoretical Physics and Astrophysics, Institute of Physics, Silesian University in Opava, Bezru\v{c}ovo n\'{a}m.13, CZ-74601 Opava, Czech Republic}

%\date{\today}
\begin{abstract} 
Buchdahl star is the most compact object without an event horizon and is an excellent candidate for a black hole mimicker. Unlike black holes, rotating Buchdahl star can be over-extremal with respect to the black hole, sustaining a larger spin. We show that it can also develop an ergosphere above the threshold spin $\beta > 1/\sqrt{2}$, which allows extraction of its rotational energy. Electromagnetic field around Buchdahl star is also expected to differ from that of black hole in both strength and topology. In this paper, we explore the energetics of Buchdahl star focusing on the magnetic Penrose process in the two magnetic field configurations, i.e., uniform and dipole. Below the threshold spin, Buchdahl star is expected to be quiet, while above the threshold it can be much more efficient than the black hole if a dipolar magnetic field is developed on its surface. 

\end{abstract}

%\pacs{}
\maketitle

\section{Introduction}
\label{introduction}
In recent years, much progress has been made in finding evidence for the existence of black holes (BH) in nature, including the detection of gravitational waves as a result of BH mergers \cite{Abbott16a,Abbott16b} by the LIGO-VIRGO collaboration, imaging supermassive black holes by the Event Horizon Telescope (EHT) \cite{Akiyama19L1,2022ApJ...930L..12E}, and measuring the orbital parameters of stars and matter around Galactic center supermassive black hole \cite{2008ApJ...689.1044G,2010RvMP...82.3121G} by GRAVITY@ESO and other facilities. However, the current precision of measurements, despite a great success that allowed one to rule out many hypothetical objects, still leaves an open window for a variety of BH alternatives, including horizonless compact objects. This makes it challenging to find the proper techniques and observational methods for distinguishing these alternatives from BH. Therefore, it is important first to identify the unique signatures of BH alternatives from both theoretical and observational perspectives. 

Among these signatures are the purely geometric characteristics of the spacetime metric, which is expected to deviate from that of the Kerr BH. However, if the deviation parameter is small, it might be observationally difficult to tell them apart. 
On the other hand, one can argue that the horizonless compact objects can have different magnetic field topologies at the event horizon scales of the corresponding BH. If the compact object has a surface of finite density capable of sustaining currents, this may cause the object to develop its own internal dipolar (or multipolar) magnetic field, which is forbidden in the case of BH by the no-hair theorem, where the magnetic field can be either monopole or external. The polarimetric observations combined with the non-thermal radiation models can already say a lot about the magnetic field orientation and strength in the environments of the BH candidates (see, e.g. \cite{EventHorizonTelescope:2021bee,EventHorizonTelescope:2021srq}). This promises to open new opportunities for distinguishing the BH from their alternatives. 

Recently, the Buchdahl star (BS) was defined by saturation of the famous Buchdahl compactness bound, $M/R \leq 4/9$, where $M$ and $R$ are the mass and radius of the compact object \cite{Dadhich22,Alho22PRD,Shaymatov23PLB,Shaymatov23JCAP}. It is universally defined by $\Phi(r) = 4/9$, where $\Phi(r)$ is the gravitational potential felt by a radially (or axially) falling particle into a static (or rotating) object. Note that BH is defined by $\Phi(r) = 1/2$ and hence BS is indeed very close to BH and is the most compact object without horizon. 
However, there is a non-trivial caveat to be acknowledged upfront that the Kerr metric describes only a rotating BH and not a non-BH rotating object. In the static case, Schwarzschild or Reissner-Nordstr\"{o}m metric describes a static object, whether a black hole or not. This is not so for the stationary axially symmetric spacetime. Kerr metric can only describe a rotating BH with spherical topology when all the multipole moments due to rotation have evaporated away. This is, however, not the case for rotating BS, which has a non-null timelike boundary. Unfortunately there exists no exact solution describing a non-BH rotating object. In the absence of this, we are forced to stick to the Kerr metric with the justification that BS is though not a BH but is very close to it as $\Phi(R) = 4/9$ instead of being 
$1/2$ for BH.

It is important to mention that neutron star with the maximum stable mass on the verge of gravitational collapse to a BH could be described in a good approximation by the Kerr metric. 
In \cite{2013MNRAS.433.1903U} a detailed analysis of the quadrupole moments of rotating neutron stars and strange stars was provided. Their study shows that for neutron stars, the exterior spacetime can be well approximated by the Kerr metric for stars near the maximum stable mass. Specifically, they find that the quantity ${QM}/{J^2}$ (where $Q$ is the quadrupole moment, $M$ is the mass, and $J$ is the angular momentum) approaches $1$ (i.e., the Kerr black hole limit)  for neutron stars at high compactness, regardless of the equation of state, indicating a close approximation to the Kerr metric. 
Similar conclusions were also made in \cite{2015PhRvD..92b3007C}, stating that the Kerr metric can serve as a good approximation for the exterior spacetime of rotating neutron stars when the compactness of the star reaches its maximum value (see, also e.g. \cite{1999ApJ...512..282L,2004MNRAS.350.1416B}). Thus, if the neutron star in the limiting compactness state can justifiably be described by the Kerr metric, so should be the case for BS. 
With this approximation acknowledged, we shall proceed with the Kerr metric for a rotating BS.

The Buchdahl bound is derived under very general conditions without reference to any specific equation of state or other fluid properties, except that the density is non-increasing outward and the interior metric matches the Schwarzschild vacuum metric at the boundary. It should also be noted that BS is the most compact object without an event horizon, while BH is the most compact object with an event horizon.

It is interesting to note that the BS is compact enough to be surrounded by the ergosphere just above its surface if the rotational parameter of the BS is sufficiently large. In this case, the ergosphere is defined as a region bounded by the static limit and the BS surface, instead of the event horizon in the case of BH. 
The existence of the ergosphere is essential for energy extraction mechanisms, as we discuss further. Since the radius of the BS is larger than the BH horizon, the ergosphere cannot go all the way to $\theta=0$, but stops at $\theta = \arccos(\sqrt{7}/3) \approx \pi/6$ for maximum spin. 
It is important to note that the extremality for BS with spin $a$ reaches at $a/M=9/8>1$, and therefore it is over-extremal relative to BH. This implies that the BS can hold a larger rotation or charge, as compared to the corresponding BH \cite{Dadhich22}. This is another reason to explore the energetics of BS, which, as we will show below, differs from that of BH.

In 1969, Penrose proposed an ingenious process \cite{Penrose:1969pc}, by which the rotational energy could be extracted from a rotating black hole. 
He pointed out that a test particle inside the ergosphere has to co-rotate with the black hole even if it has negative angular momentum. The remarkable feature of this property is that a particle with negative angular momentum in the ergosphere could also have negative energy relative to infinity. It is envisaged that if a particle falls from infinity with a certain energy and splits into two fragments in the ergosphere, one of which attains negative energy and falls into the hole, the other would then come out with an energy greater than that of the incident particle. This is how the rotational energy of a black hole could be mined out by the Penrose process.

This is a remarkable fully general relativistic process that gave rise to considerable excitement for it could in principle power the then recently discovered high energy exotic objects like quasars and active galactic nuclei (AGNs). However, the excitement was short-lived as detailed calculations by Bardeen et al. \cite{Bardeen72} and independently by Wald \cite{Wald74ApJ} showed that for the process to be operative, the relative velocity between the two splitting fragments has to be greater than $1/2\,c$. There could be no conceivable astrophysical phenomena that could accelerate the fragments to such a high speed instantaneously. The process was very novel and elegant,  entirely driven by the geometry of spacetime, but it could not be astrophysically viable. Some variants involving photons and collisions of particles were considered, see \cite{Dadhich12MPP,Tursunov:2019oiq,2021Univ....7..416S}, but none could make the process astrophysically significant. 

Independent of this, Blandford and Znajek proposed a process of energy extraction from rotating black holes that involves electrodynamics of the force-free plasma \cite{Blandford1977}. It is driven by the twisting of the magnetic field lines due to the frame dragging effect of the Kerr geometry builds up a potential difference between the pole and the equator, and discharging of that drives rotational energy out.  
Following that, the Penrose process was generalized to the case with the presence of a magnetic field around a rotating black hole by Wagh et al. \cite{Wagh85ApJ}, in which the energy required for the particle to attain negative energy could now come from an electromagnetic field totally circumventing the formidable velocity limit. This marked the revival of the Pensose process as the magnetic Penrose process (MPP), and, more importantly, it can be highly efficient with efficiency even exceeding 100\% \cite{Bhat85,Parthasarathy86,Tursunov:2019oiq,Shaymatov22b}.

More recently, it has been shown \cite{Dadhich18} that MPP is a general energy extraction process, which encompasses the BZ mechanism for a moderately high magnetic field. Most importantly, it has been successfully employed in modeling as a source of ultra-high energy cosmic rays (UHECR) \cite{Tursunov20ApJ,2022Symm...14..482T}. There exists an extensive body of work covering a wide variety of situations \cite{Wagh89,Morozova14,Alic12ApJ,Moesta12ApJ,McKinney07,Tursunov21:EP,Kolos:2020gdc} that address the electromagnetic field's impact on the energy extraction from black holes in connection with accretion disks and jets.  
It has also been shown that energy extraction process can be strongly affected by the gravitomagnetic monopole charge~\cite{Abdujabbarov11}.

In this paper, our aim is to explore the conditions of rotational energy extraction from rotating BS involving a magnetic field of two different topologies. First, we study the energy extraction from BS that is immersed into external asymptotically uniform magnetic field, i.e in a setup similar to previous studies related to BH. Second, we assume that BS develops its own magnetic field of a dipolar topology, which cannot be applied in the case of BH, and explore the MPP in this case. The paper is organized as follows. In Sec.~\ref{Sec:metric} we introduce the conditions for the existence of the ergosphere around BS and briefly recall some of the relevant properties of the Kerr metric in the context of BS. In Sec.~\ref{Sec:GeometricPP} we study the geometric Penrose process for BS and find the maximum efficiency of this process. In Sec.~\ref{Sec:U&DMF} we discuss the electromagnetic fields of two configurations in Kerr geometry: \textit{uniform} and \textit{dipole}.  
In Sec.~\ref{Sec:Penrose} we formulate MPP, calculate its efficiency for two magnetic field configurations, and compare the results with that of the BH case. 
In Sec.~\ref{Sec:conclusion} we discuss astrophysical implications of our results and give our concluding remarks.  
Throughout the manuscript we utilize the metric signature $(-,+,+,+)$ and use the system of geometric units, in which $G=c=1$.

\section{Ergosphere of  Buchdahl star  }\label{Sec:metric}

\begin{figure*}
%\begin{center}
  \includegraphics[width=\textwidth]{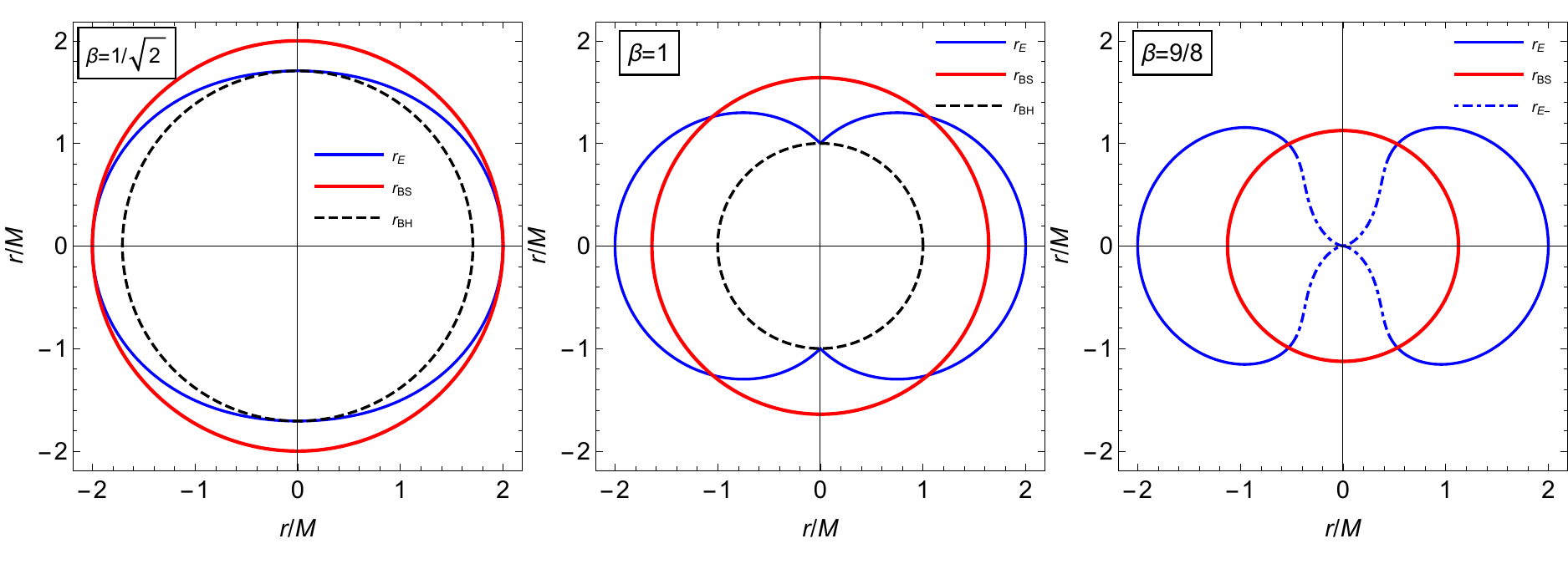}
%  &  \includegraphics[scale=0.6]{vel2.pdf}
	\caption{\label{fig:fig1} 
Sizes and relative positions of the ergosphere $r_E$ (blue), surface of the Buchdahl star $r_{BS}$ (red), and the event horizon of a corresponding black hole $r_{BH} \equiv r_{+}$ (black dashed) for different values of the spin parameter $\beta$. From left to right is shown the ergosphere  for $\beta = 1/\sqrt{2}, \, 1, \, 9/8$, respectively. Note that the left panel shows the threshold case of absence of the ergosphere as $r_E < r_{BS}$ while it exists for the middle and right panels but does not extend all the way to the axis, $\theta=0$. The right panel refers to the extremal BS, which is over-extremal for BH.}  
%\end{center}
\end{figure*}

As mentioned before we use the Kerr metric, which truly describes a rotating BH and not a non-BH rotating object, for a rotating BS as an approximation. It is a good reasonable approximation for the reason that it is quite close to BH object, as shown by the defining condition of the potential  $\Phi(r)=4/9$ for BS as against $1/2$ for BH. With this assumption, we write the Kerr metric for the Buchdahl star in the usual Boyer-Lindquist coordinates and that reads as 
\begin{eqnarray}\label{eq:metric}
ds^2 &=&-\left(\frac{\Delta-a^2\sin^2\theta}{\Sigma}\right)dt^2 +\frac{\Sigma}{\Delta}dr^2+\Sigma d\theta^2\nonumber\\
&-&\frac{2a\sin^2\theta(r^2+a^2-\Delta)}{\Sigma}dtd\phi \nonumber\\
&+&\frac{(r^2+a^2)^2-\Delta a^2\sin^2\theta}{\Sigma}\sin^2\theta
d\phi^2 \, ,
\end{eqnarray}
where $\Sigma=r^2+a^2\cos^2\theta$ and $\Delta=r^2+a^2-2Mr$ with the two parameters, $a$ and $M$ denoting respectively spin and mass of the star respectively.  

In this spacetime gravitational potential felt by an axially falling timelike particle would be given by 
\begin{eqnarray}
\Phi(R) = \frac{M\,r}{(r^2+a^2)}=\frac{M/r}{1+\beta^2 (M/r)^2}\, ,
\end{eqnarray} 
where $\beta^2=a^2/M^2$. 
Thus BS is defined by \cite{Dadhich22,Dadhich20:JCAP}
 \begin{eqnarray}
\Phi(r) = \frac{M/r}{1+\beta^2 (M/r)^2} = 4/9\, ,
\end{eqnarray}
that solves to give 
\begin{eqnarray}
\frac{M}{r}
=\frac{8/9}{1+ \sqrt{1-(8/9)^2{\beta^2}}} \, ,
\end{eqnarray}
i.e, Buchdahl star surface radius is 
\begin{eqnarray} \label{eq:rBS}
r_{BS} =\frac{9M}{8}\left(1+\sqrt{1-(8/9)^2{\beta^2}}\right)\, .
\end{eqnarray}
Similarly, for black hole the horizon is given by $r_{+} = M(1+ \sqrt{1-\beta^2})$.

The Kerr metric admits two Killing vectors, i.e. $\xi^{\alpha}_{(t)}=(\partial/\partial t)^{\alpha}$ and
$\xi^{\alpha}_{(\varphi)}=(\partial/\partial \phi)^{\alpha}$ that are responsible for stationarity and axial symmetry
of spacetime, thus leading to the conserved quantities that respectively refer to energy and angular momentum of a test particle. There also exists the static surface defined by the timelike Killing vector $\xi^{\mu}_{(t)} = \partial/\partial t$ turning null, i.e., $g_{tt}= 0$ which is given by 
\begin{eqnarray}
r_{st}(\theta)=M\left(1+\sqrt{1-\beta^2\cos^2\theta}\right)\, .
\end{eqnarray}

That is, no particle can remain static at a fixed point on or below this surface. The region between static surface $r_{st}$ and the horizon $r_{+}$ defines the ergosphere where $E = -p.\partial/\partial t$ may turn negative; i.e, a timelike particle may have negative energy relative to an observer at infinity. This is the key property responsible for energy extraction by the Penrose process. In the case of Buchdahl star the ergosphere is the region lying between $r_{BS}$ and $r_{st}$ as shown in Fig.~\ref{fig:fig1}. Of course the width of ergosphere is smaller than that for BH because $r_{BS} > r_{+}$. However note that the extremality for BS is at $\beta=9/8 >1$ and hence extremal BS is over-extremal relative to BH. The ergosphere has maximum width at extremality, which for BH is $r_{st} - r_{+}(\beta=1) = 2M-M = M$, while that for BS $r_{st} - r_{BS}(\beta=9/8) = 2M - 9/8\,M = 7/8\, M$.  

\begin{figure}
%\begin{center}
  \includegraphics[scale=0.65]{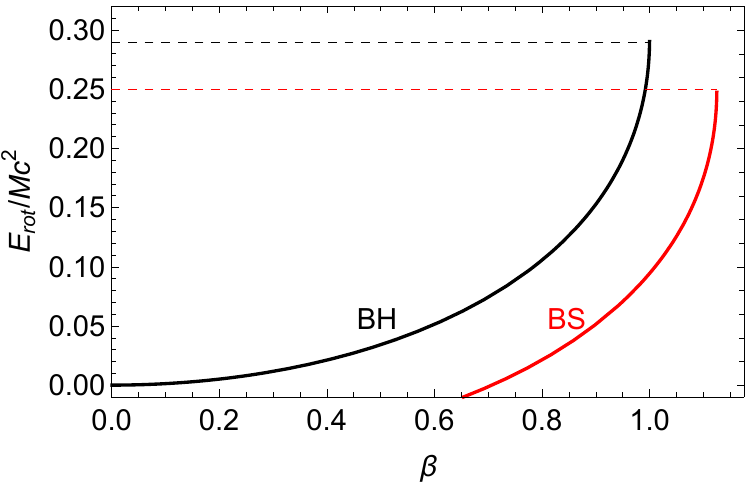}\hspace{0.0cm}
	\caption{\label{fig:erot} Extractable fraction of total energy of black hole (black) and Buchdahl star (red) plotted against $\beta$. The maximal values are indicated by the dashed lines.  
 }
%\end{center}
%
\end{figure}

The ergosphere is defined by $r_{st} -r_{BS}$ that vanishes for $\beta = \beta_* = 1/\sqrt2 \sim 0.71$. In Fig.~\ref{fig:fig1} it is shown that the ergosphere for BS can be above its surface only for $\beta>1/\sqrt{2} \sim 0.71$. This implies that in contrast to the BH case, in the case of BS, the rotational energy is extractable only for rapidly rotating BS with $\beta>1/\sqrt{2} \sim 0.71$. 
Indeed, one can find the extractable rotational energy of the BH/BS as follows
\begin{eqnarray}\label{Eq:mass_irr}
E_{rot} = M - M_{ir}\,\, , M_{ir}^2 = \frac{1}{2} Mr_*\, ,
\end{eqnarray}
where $r_* = r_+$ for a black hole and $r_* = r_{BS}$ for Buchdahl star. This relation is obtained by comparing the area of rotating BH/BS with the Schwarzschild BH/BS %($r^2=4M_{ir}^2$, 
defining the irreducible mass $M_{ir}$. With rotation, it is possible to integrate for area in closed form at the event horizon for BH while it is not so for at the Buchdahl radius for BS. In the latter case, as an approximation we evaluate $r^2+a^2$, as is the case for horizon, at $r_{BS}$. Evaluating Eq.~(\ref{Eq:mass_irr}) for extremal spins ($\beta=1$ for BH and $\beta=9/8$ for BS) we obtain $E_{rot}^{max}=0.29 M$ and $E_{rot}^{max}=0.25 M$, respectively.  In Fig.~\ref{fig:erot} we have plotted the extractable rotational energy for BH and BS against the rotation parameter, $\beta$. 
Note that extractable energy goes to zero for the threshold $\beta_*$. 

Since we have taken the same Kerr metric for BS, the spacetime around BS is equivalent to that of the BH above the BS surface. This implies that the geodesics of test particles and photons will be the same in that region and, therefore, cannot be used to effectively distinguish the two objects from each other. However, BS with the spin parameter less than the critical, i.e. $\beta\leq 1/\sqrt2$, would be very inefficient in a sense of relativistic jet launching \cite{Blandford1977}  and particle acceleration mechanisms \cite{Tursunov20ApJ}. For such subthreshold spins BS has no extractable/convertable rotational energy to power high energy jets and particle acceleration.

Note that in Eq. (\ref{Eq:mass_irr}), we evaluate for area $r^2+a^2$ at $r_{BS}$, and
for the ergosphere, $r_{st} - r_{BS}$. Strictly, these expressions
are valid only at the horizon having spherical topology. Off the horizon, the timelike BS boundary will not have spherical topology, however, we assume it to be so as an approximation. Remarkably, both of these calculations yield the same critical value for $\beta$. 

The question arises what happens to the residual rotation and its energy, could it not be extracted through the magnetic Penrose process? The magnetic Penrose process facilitates efficient extraction of rotational energy but by itself cannot extract rotational energy. That could only be extracted when there exists ergoregion. In the absence of ergoregion, rotational energy cannot be extracted. A rotating BS would therefore always end up with a residual rotation and its inextractable energy.

\section{GEOMETRIC PENROSE PROCESS}\label{Sec:GeometricPP} 

The original Penrose process of energy extraction from a rotating black hole utilized the ingenious geometric property of the existence of negative energy states near the horizon in the ergosphere. No particle can remain static fixed at a point below the static surface, and it has to co-rotate even if it has negative angular momentum relative to BH. It is this property that allows particles to have negative energy, $E=-\xi^{\alpha}_{(t)}U_\alpha$, where $\xi^{\alpha}_{(t)}=(\partial/\partial t)^{\alpha}$ is the timelike Killing vector, which turns spacelike in the ergosphere. That is why a particle's energy $E$ could be negative relative to the observer at infinity. 

In the Penrose process, it is envisaged that a particle of energy $E_1$ splits into two fragments having energy $E_2$ and $E_3$ in the ergosphere, one of which attains negative energy, say $E_2<0$ and the other comes out with energy $E_3>E_1$, i.e. greater than that of the incident particle. This is how the rotational energy of BH 
could be extracted out. However, the critical question for the process to work is, what does it take one of the fragments to ride on negative energy orbits in the ergosphere. The question was answered independently in \cite{Bardeen72, Wald74ApJ}. 
We shall briefly recall the derivation of the Wald inequality \cite{Wagh85ApJ} for the threshold relative velocity between the two fragments. 

We evaluate the invariant $E_2 = -U.\xi$ in the Killing static frame where $\xi=(1,0,0,0)$ and in the rest frame of particle 1 where $\xi=(\xi_t, \xi_a)$. Thus we write 
\begin{equation}
E_2=-U_2.\xi = \gamma(\xi_t + v_a\xi^a)=\gamma(E_1 + v\,\xi\, cos\psi)\, , 
\end{equation}
where $\gamma=(1-v^2)^{-1/2}$, and we have used $E_1=-U_1.\xi=-\xi_t$, $\xi\, \xi = g_{tt}=-E_1^2 + \xi_a\xi^a$, and $\psi$ is the angle between two 3-vectors $\xi$ and $v$.

\begin{figure}
\begin{center}
\begin{tabular}{c}
  \includegraphics[scale=0.9]{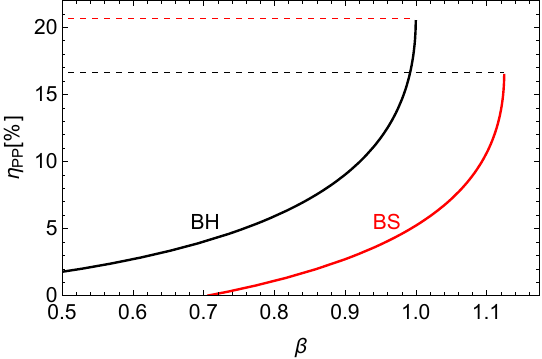}\hspace{0.0cm}
  %\includegraphics[scale=0.55]{veff_en2.pdf}\hspace{0cm}
%  &  \includegraphics[scale=0.6]{vel2.pdf}
\end{tabular}
	\caption{\label{fig:geoPP} {The efficiency of energy extraction from Kerr BH (black) and BS (red) is plotted against the dimensionless spin parameter $\beta$. The maximal values are shown by the dashed lines. }}
\end{center}
\end{figure}

Since $-1\leq cos\psi \leq1$, we write 
\begin{eqnarray}
1-v\sqrt{1+g_{tt}/E_1^2}\leq E_2/E_1\leq 1+v\sqrt{1+g_{tt}/E_1^2}\, .
\end{eqnarray}
For $E_2<0$, we then obtain the Wald inequality 
\begin{equation}
v \geq 1/\sqrt{1+g_{tt}/E_1^2}\, .
\end{equation}
This is for extremal BH where $g_{tt} =1$ gives the well-known bound $v\geq 1/2$ \cite{Wagh85}. For obtaining the velocity threshold in the Buchdahl star limit, we evaluate the above inequality for an extremal case with $\beta=9/8$ at $r_{BS}/M=9/8$ and $E_1^2=1/3$, the maximum energy for a stable circular orbit, and so we obtain  
\begin{eqnarray}
v \geq \sqrt{0.3} \approx  0.547\, .
\end{eqnarray}
This condition is even stronger than for BH, which is $v \geq 0.5$. Therefore, the geometric Penrose process is not viable astrophysically for both BH and BS.  However, we can still define the efficiency of the energy extraction as 
\begin{eqnarray}
\eta= \frac{\mbox{gain in energy}}{\mbox{input energy}}=\frac{\vert E_2\vert}{E_1}=\frac{E_3-E_1}{E_1}\, . 
\end{eqnarray}
In Fig.~\ref{fig:geoPP} we compare the maximum efficiency of purely geometric PP for Kerr BH and BS. One can see that the BH is more efficient than the BS for the same spin. The maximum value of efficiency in the Kerr BH case is $\eta=20.7$, while for BS the same is $16.7$. It can also be seen that the rotational energy of the BS is not extractable at all until the threshold value of $\beta \geq 1/\sqrt{2}$ is reached. 

However, all this gets circumvented when electromagnetic interaction is involved, as we explore in the next section. In that case, we write 
\begin{eqnarray}
E_2= \gamma\left(E_1-v\sqrt{E_1^2+g_{tt}}\right) - q_2 \Phi(r) \leq 0
\end{eqnarray}
where $\Phi(r) = A_{\alpha}\xi^{\alpha}$ is the electromagnetic potential. This will hold good even when $v=0$ because of the electromagnetic contribution. That is energy required for particles to attain negative energy could now come from electromagnetic interaction totally removing any threshold on relative velocity between the fragments. This is how Penrose process in its magnetic version got revived as the most attractive powering mechanism for central engine of high energy astrophysical objects. In the next sections we first discuss plausible magnetic field configurations, which can arise around rotating BS,  followed by the magnetic Penrose process and its high efficiency. 

\section{ELECTROMAGNETIC FIELD AROUND ROTATING BUCHDAHL STAR }\label{Sec:U&DMF}

The magnetic field strength and structure can be different for BH and BS.  
It is well known that isolated BH cannot have its own magnetic field, except magnetic monopole, leading to the no-hair theorem for BH. However, there are external factors that become increasingly important for a magnetic field to exist in the surrounding environment of black holes \cite{Stuchlik20}. The magnetic field can be caused e.g. due to the presence of accretion disk of magnetized plasma. The strength of the magnetic field then will depend on the density of accreting matter and other parameters of the accretion disk. BH can also be immersed into external magnetic field of a companion neutron star. In that case, if the distance from the source is far enough, the magnetic field configuration at the event horizon scale of the BH can be well approximated to the uniform solution \cite{Wald74}. If the BH has a relativistic jet, which requires a large scale magnetic field to be present, then the field lines developed at the event horizon scale can be of a paraboloidal shape \cite{2023EPJC...83..323K}. A typical magnetic field strength measured on the Schwarzschild radius scale of many supermassive BH in AGN is around the order of $10^4$G \cite{Daly:APJ:2019:,Baczko16}, although for some sources an order of magnitude weaker magnetic fields have also been reported \cite{EventHorizonTelescope:2021bee,EventHorizonTelescope:2021srq}.

The same lines of argument for the presence of a magnetic field around BH can also be applied to BS. However, in addition and in contrast to BH, BS can potentially also develop its own magnetic field if electric currents are sustained on its surface. These fields for rotating BS are likely to be dipolar in shape, which cannot be the case for BH. Given the compactness of the BS, the own magnetic field strength, if developed, is expected to be relatively strong, perhaps comparable to or even exceeding the magnetic field strengths of neutron stars. Therefore, in the following, we test the electromagnetic extraction of rotational energy of BS in two different magnetic field configurations: uniform and dipole.

Further, we assume that magnetic fields under consideration are weak for any alteration of the background spacetime geometry. This implies that the magnetic field can always be taken as a test field in curved background spacetime, which is well justified astrophysically (see, e.g. discussion in Ref.~\cite{Tursunov16}). We also assume the magnetic field shares the axial-symmetry of the BS. While in BH it is a natural assumption, it is not necessarily so for BS, especially if BS has its own dipole magnetic field. However, for simplicity, we will follow the assumption of sharing axial symmetry of the field and the BS spin. Then, for magnetic field along the axis of symmetry, one can write 
\begin{equation}
A^{\alpha}=C_1 \xi^{\alpha}_{(t)}+ C_2 \xi^{\alpha}_{(\varphi)} ,
\end{equation}
with stationary $\xi^{\alpha}_{(t)}=(\partial/\partial t)^{\alpha}$ and
axial $\xi^{\alpha}_{(\varphi)}=(\partial/\partial \phi)^{\alpha}$ Killing vectors. Note that $C_1$ and $C_2$ are integration constants that specify the nature of field. Without going into details of specific solutions, which can be found in the literature,   
we will only focus on the covariant time components of the four-vector potentials of the considered electromagnetic fields, which will be needed further for the calculation of efficiencies of the MPP. 

For a uniform magnetic field, the covariant time component of the vector potential $A_{\mu}^{\rm U}$ is given by \cite{Wald74}
\begin{eqnarray}\label{Eq:4-pot_t}
A_t^{\rm U} &=& a B \left[\frac{M r}{\Sigma}(1+\cos^2\theta)-1\right]\, , 
\end{eqnarray}
where $B$ is the strength of external asymptotically uniform magnetic field at infinity. 
For the dipole magnetic field, we employ the model of the magnetic field generated by an electric current-carrying loop placed at the surface of the rotating BS (see details in Refs.~\cite{Chitre75PRD,Petterson75PRD}). The electromagnetic field can then be written as follows:  
\begin{eqnarray}\label{Eq:4-pot_t1}
A_t^{\rm D} &=& 2\pi I \left(\frac{r^2-2Mr+a^2}{r^2+a^2+2Ma^2/r}\right)^{1/2}\frac{a}{r}\, ,
\end{eqnarray}
where $I$ is an electric current strength generated by the loop. All details and electromagnetic field structure were discussed in Refs.~\cite{Chitre75PRD,Petterson75PRD}.  

The presence of a magnetic field would influence the motion of charged particles through the Lorentz force, which would be significant even for a very weak field.  There exists a vast literature on addressing the impact of magnetic field on various astrophysical processes and on testing the background geometry in close vicinity of black holes~\cite[see, e.g.][]{Aliev02,Frolov10,Jamil15,Tursunov16,Shaymatov22a,Hussain17,Stuchlik20,Shaymatov21pdu,2023EPJC...83..323K,Shaymatov22c}. Below, we briefly repeat the derivation of the leading dynamical equations for a test charged particle in the most general form, which will be useful for the formulation of the MPP. 

The dynamics of a charged particle in the background electromagnetic field $A^\mu$ and curved spacetime $g_{\mu\nu}$ is governed by the Hamiltonian~\cite{Misner73}
\begin{eqnarray}
H=\frac{1}{2}g^{\mu\nu} \left(\pi_{\mu}-qA_{\mu}\right)\left(\pi_{\nu}-qA_{\nu}\right)\, ,
\end{eqnarray}
where $\pi_{\mu}$ refers to the canonical momentum and $A_{\mu}$ is the electromagnetic four-vector potential given by either of Eqs.~(\ref{Eq:4-pot_t}) and (\ref{Eq:4-pot_t1}). 
The relation between the four-momentum and the canonical momentum of the charged particle is given by $p^{\mu}=g^{\mu\nu}\left(\pi_{\nu}-qA_{\nu}\right)$. Further, one can write the equation for the time-like particle motion as
\begin{eqnarray} 
\label{Eq:eqh1}
  \frac{dx^\alpha}{d\lambda} = \frac{\partial H}{\partial \pi_\alpha}\,   \mbox{~~and~~}
  \frac{d\pi_\alpha}{d\lambda} = - \frac{\partial H}{\partial x^\alpha}\, , 
\end{eqnarray}
with affine parameter $\lambda=\tau/m$,  where $\tau$ is the proper time of the particle. In the equatorial plane, $\theta=\pi/2$, the motion of the particle will be bounded by the energetic effective potential, which can be written by  
\begin{eqnarray}
V_{eff}(r)&=&-\frac{q}{m}A_{t}+\omega \left(
\mathcal{L}-\frac{q}{m}A_{\phi}\right)\nonumber\\&+&\sqrt{\frac{\Delta}{g_{\phi\phi}}\left(\frac{
\left(\mathcal{L}-\frac{q}{m}A_{\phi}\right)^2}{g_{\phi\phi}}+1\right)}\, ,
\end{eqnarray}
where $\mathcal{L}=L/m$ is the angular momentum per mass of the particle and $\omega=-g_{t\phi}/g_{\phi\phi}$ is the angular frame dragging velocity of zero angular momentum particle. Since the motion of a charged particle always has a barrier at the equatorial plane \cite{Tursunov16}, we consider a special case of motion in the circular orbit with constant $r$. Then we have ${\bf u}\sim {\bf \xi}_{(t)}+\Omega {\bf \xi}_{(\phi)}$, where $\Omega=d\phi/dt=u^{\phi}/u^{t}$ is the angular velocity of the particle measured by an observer at infinity. For timelike particle, the angular velocity is bounded \cite{Parthasarathy86,Wagh:1985vuj} as $\Omega_{-}<\Omega<\Omega_{+}$ where  \begin{eqnarray}
  \Omega_{\pm} =\frac{-g_{t\phi}\pm \sqrt{(g_{t\phi})^{2}-g_{tt}g_{\phi\phi}}}{g_{\phi\phi}}\, .
\end{eqnarray}
is the angular velocity of outer and inner photon circular orbit. 
Note that at the static surface, $g_{tt}=0$, $\Omega_{+}=0$ and $\Omega_{-}=2\omega$. For the timelike circular orbit, we write $\pi_{\pm}=p^{t}(1,0,0,\Omega_{\pm})$ giving the equation,  
\begin{eqnarray}\label{Eq:W0}
\left(g_{\phi\phi}\pi_t^2+  g_{t\phi}^2\right)\Omega^2&+&2g_{t\phi}\left(\pi_t^2+g_{tt}\right)\Omega\nonumber\\&+& g_{tt}\left(\pi_t^2+g_{tt}\right)=0\, ,
\end{eqnarray}
where $\pi_t=-\left({E}+qA_{t}\right)/m$. The above equation solves to give angular velocity of the particle \cite{Parthasarathy86,Nozawa05}
\begin{eqnarray}\label{Eq:an_veloffallin}
\Omega=\frac{-g_{t\phi}\left(\pi_t^2+g_{tt}\right)+\sqrt{\left(\pi_t^2+g_{tt}\right)\left(g_{t\phi}^2-g_{tt}g_{\phi\phi}\right)\pi_t^2}}{g_{\phi\phi}\pi_t^2+g_{t\phi}^2}\, .\nonumber\\
\end{eqnarray}  

In the next section, we will focus on the MPP and its efficiency, based on the analysis of the charged particle motion.% from the literature.

\section{MAGNETIC PENROSE PROCESS }\label{Sec:Penrose}

\begin{figure*}
\begin{tabular}{c c }
  \includegraphics[scale=0.52]{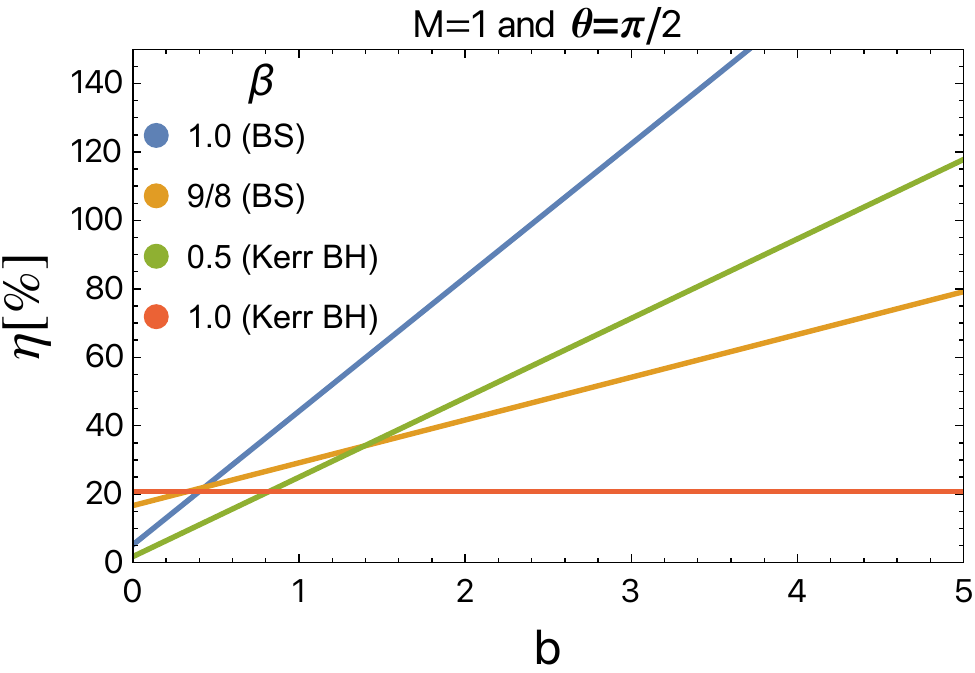}\hspace{-0.0cm}
  &\includegraphics[scale=0.565]{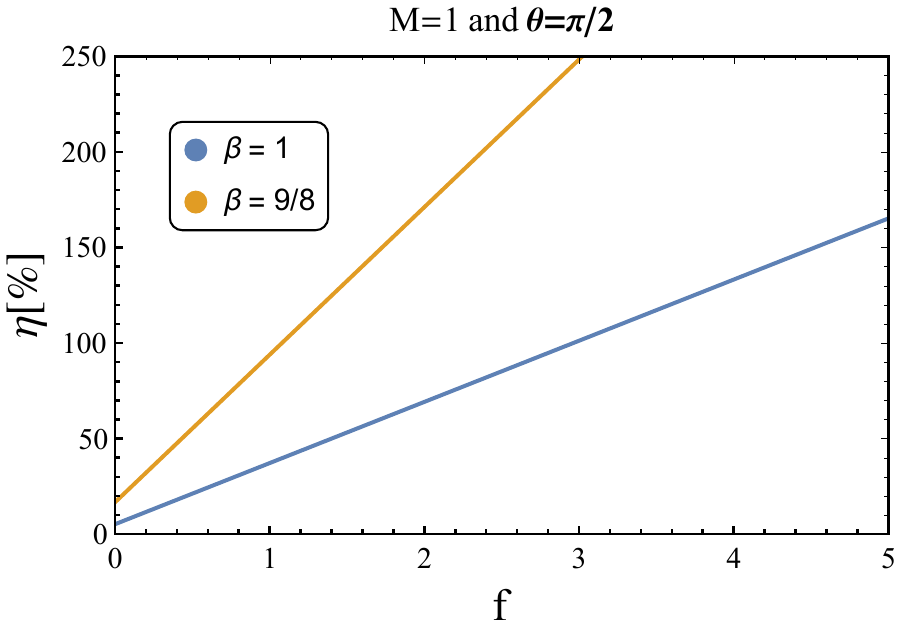}\\ \includegraphics[scale=0.52]{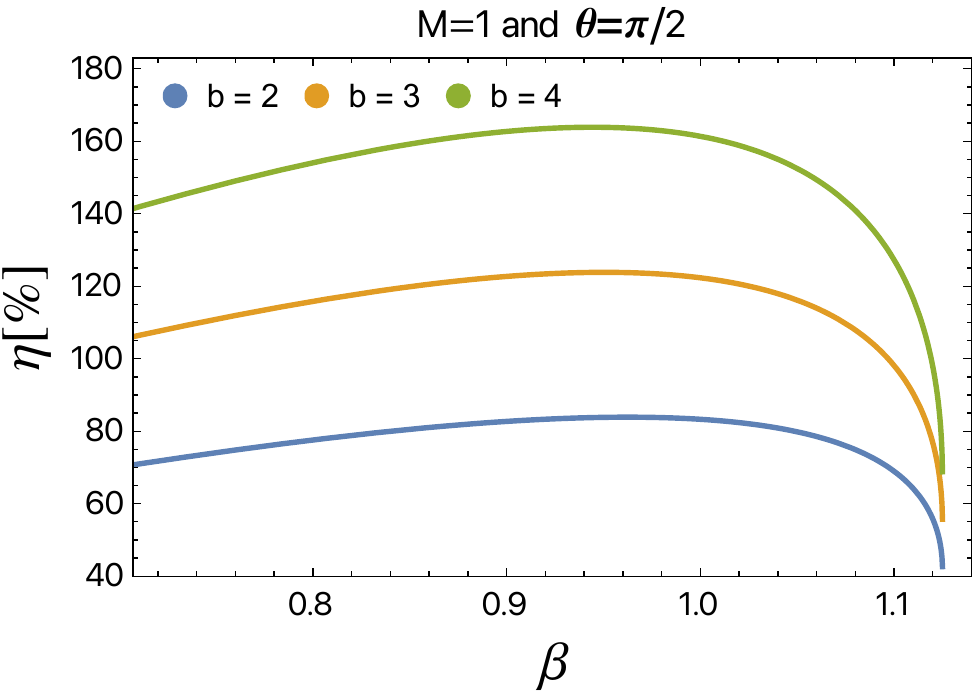}\hspace{-0.0cm} &\includegraphics[scale=0.565]{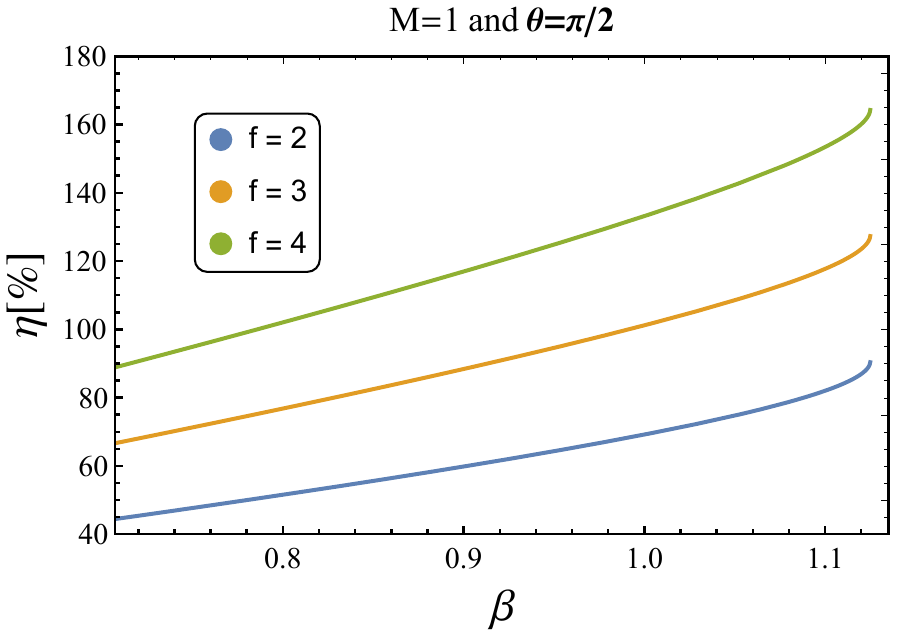}
\end{tabular}
\caption{\label{fig:mpp} Efficiency of energy extraction from BS for uniform (left column) and dipole (right column) configurations. 
}
\end{figure*}
We now turn to MPP in which it is envisaged that a neutral particle splits into two charged fragments in the ergosphere threaded with asymptotically uniform magnetic field. We have already seen that ergosphere does exist for BS, and so a particle with suitable parameters can attain negative energy state in there. Let the incident particle has energy $E_{1}$, which splits into two fragments having energy $E_{2}$ and $E_{3}$ and charge $-q$ and $+q$ respectively. Let $E_{2}<0$ fall into BS, then by conservation of energy $E_{3}= E_{1}-E_{2} > E_{1}$ would come out with the larger energy than the incident $E_{1}$. The presence of magnetic field circumvents the formidable velocity threshold, $v\geq 1/2$ of the original geomtric PP and thereby enormously increases the efficiency of energy extraction. The extracted energy is the rotational energy of the BS, which can be up to 25\% of the total BS mass energy. Thus, we write 
\begin{eqnarray}\label{Eq:con_laws}
\begin{cases}
E_1=E_{2}+E_{3}\, \\
L_1=L_{2}+L_{3}\,  
\end{cases}
\mbox{~~~and~~~}
\begin{cases}
m_1=m_{2}+m_{3}\, \\
q_1=q_{2}+q_{3}\, , 
\end{cases}
\end{eqnarray}
and 
\begin{eqnarray}\label{Eq:con_law}
m_1u_1^{\mu}&=& m_2u_2^{\mu}+m_3u_3^{\mu}\, .
\end{eqnarray}
Note that $u^{\phi}=\Omega\,u^{t}=-\Omega A/B$ and $u^{t}=\pi_{t}/\left(g_{tt}+\Omega g_{t\phi}\right)$. Now Eq.~(\ref{Eq:con_law}) reads as follows:  
\begin{eqnarray}
\Omega_1m_1A_{1}B_2B_3=\Omega_2m_2A_{2}B_3B_1+\Omega_3m_3A_{3}B_2B_1\, ,
\end{eqnarray}
with $A_{i}=\left({E_{i}} +qA_{t}\right)/m_{i}$\footnote{$A_i$ here should not be confused with the spatial components of the electromagnetic four-potential $A_\mu$} and $B_{i}=g_{tt}+\Omega_{i}\,g_{t\phi}$. From the above equation, we write \cite{Tursunov:2019oiq} 
\begin{eqnarray}
\frac{E_3+q_3A_{t}}{E_1+q_1A_{t}}=\left(\frac{\Omega_1B_2-\Omega_2B_1}{\Omega_3B_2-\Omega_2B_3}\right)\frac{B_3}{B_1}\, ,
\end{eqnarray}
which yields energy of the escaping fragment  
\begin{eqnarray}\label{Eq:E3}
E_3=\chi\left(E_1+q_1A_{t}\right)-q_3A_{t}\,, 
\end{eqnarray}
where we define 
\begin{eqnarray}\label{Eq:chi}
\chi=\left(\frac{\Omega_1-\Omega_2}{\Omega_3-\Omega_2}\right)\frac{B_3}{B_1}
\, ,
\end{eqnarray}
where $\Omega_1= \Omega\, , \mbox{~~} \Omega_2=\Omega_{-}\, \mbox{~~and~~} \Omega_3=\Omega_{+}$.

Let us now define the efficiency of the energy extraction in the MPP as follows 
\begin{eqnarray}
\eta= \frac{\mbox{gain in energy}}{\mbox{input energy}}=\frac{\vert E_2\vert}{E_1}=\frac{E_3-E_1}{E_1}\, . 
\end{eqnarray}
From Eqs.~(\ref{Eq:E3}) and (\ref{Eq:chi}), the explicit form of $\eta$ reads as 
\begin{eqnarray}
\eta= \chi-1+\frac{q_3A_t}{m_1\pi_{t1}+q_1A_t}-\frac{ q_1A_t}{m_1\pi_{t1}+q_1A_t}\,\chi\, .
\end{eqnarray}
It can alternatively be written as 
\begin{eqnarray}
\eta= \left(\frac{\Omega-\Omega_{-}}{\Omega_{+}-\Omega_{-}}\right)\left(\frac{g_{tt}+\Omega_{+}g_{t\phi}}{g_{tt}+\Omega\,g_{t\phi}}\right)-1-\frac{q_3A_t}{E_1}\, ,
\end{eqnarray}
where we have taken $q_2=-q=-q_3$. 

The efficiency of energy extraction increases by getting the splitting point closer to the surface of the BS, $r_{BS}$, where it reaches the maximum.  
For the uniform magnetic field configuration, using Eq.~(\ref{Eq:4-pot_t}) we get the maximum efficiency of the MPP as follows 
\begin{eqnarray}\label{Eq:q}    
\eta_{\rm max}^{\rm U}&=& 
\frac{1}{2} \left(\sqrt{\frac{16/9}{1+\sqrt{1-(8/9)^2\beta^2}}}-1\right) \nonumber\\&+&\beta\, {b}M\left(1-\frac{8/9}{1+\sqrt{1-(8/9)^2\beta^2}}\right)\, ,
\end{eqnarray}
where we introduced the dimensionless interaction parameter $b$ characterizing the relative ratio between the Lorentz and gravitational forces 
\begin{eqnarray}\label{Eq:mag_par}
b=\frac{qBGM}{mc^4}\, .
\end{eqnarray} 
Here $B$ indicates the strength of the uniform external magnetic field. Note here that Eq.~(\ref{Eq:q}) consists of two parts, where the first part gives the efficiency of the geometric PP (for neutral particles), while the second is the magnetic part (charged particles); i.e. $\eta = \eta\vert_{geometric, q=0} + \eta\vert_{magnetic, q\neq0}$. The above expression for the Kerr BH could be obtained by writing $8/9\to1$ everywhere, which will yield the maximum value of $20.7\%$ when $b=0$, and the magnetic part goes to zero as $\beta\to1$. 

For the dipole magnetic field configuration generated by the current loop at the BS surface (\ref{Eq:4-pot_t1}), the maximum efficiency is given by 
\begin{eqnarray}\label{Eq:I}    
\eta_{\rm max}^{\rm D}&=& 
\frac{1}{2} \left(\sqrt{\frac{16/9}{1+\sqrt{1-(8/9)^2\beta^2}}}-1\right) \nonumber\\&+&\pi \,\beta\, {f}
 \frac{\left(153-9\sqrt{1-(8/9)^2\beta^2}\right)}{1458+4\beta^2}\, ,
\end{eqnarray}
with the interaction parameter $f$ given by 
\begin{eqnarray}\label{Eq:mag_par_dip}
f=\frac{qI}{mc^2}\, .
\end{eqnarray} 
Here $I$ indicates the strength of the electric current of the loop at the BS surface. Similarly to the uniform magnetic field case, Eq.~(\ref{Eq:I}) also consists of two parts, where the first part gives the efficiency of the geometric PP, while the second is the contribution due to interaction of charged particles with the dipole magnetic field of the BS. The dependencies of the efficiencies of MPP for uniform and dipole magnetic field configurations are shown in Fig.~\ref{fig:mpp}.

If the splitting point of an incident particle occurs at $r\geq r_{\rm BS}$, the efficiency of the MPP for BS with the spin parameter in the range $1/\sqrt{2} < \beta < 1$, will be essentially the same as in the case of BH of the same mass and spin, thus fully mimicking BH. For smaller spins, i.e. for $\beta \leq 1/\sqrt{2}$, the MPP is inoperative, as there is no extractable energy of BS for such spin values. Contrary to the BH, BS can be over-extremal by having the spin in the range $1 < \beta \leq 9/8$. Since the efficiency of MPP increases with increasing spin of the BS, one may think that overextremal BS is more efficient than the rapidly rotating BH of the same mass. It is so, if the splitting point of an incident particle is chosen to be above the Buchdahl radius $r\geq r_{\rm BS}$. However, since BH allows the splitting point to occur below the Buchdahl radius,  at $r_{+} \leq r < r_{\rm BS}$, the BH with a moderate spin can eventually be more efficient than the overextremal BS, although the difference is small due to the dominant effect of the magnetic field over gravity in astrophysical situations \cite{Tursunov:2019oiq}.

\section{Discussion and Conclusion}
\label{Sec:conclusion}

One of the most promising sources of energy for powering high energy astrophysical objects is the rotational energy of black hole (BH). To that, in this paper, we have added one more interesting candidate, rotating Buchdahl star (BS). While for BH, the rotational fraction of the total energy can be as high as $E_{rot}^{max} = 0.29 M$, for BS the same is very close, being equal to $E_{rot}^{max} = 0.25 M$. Since above the Buchdahl radius (\ref{eq:rBS}), the spacetime around BS is equivalent to that of the BH, the geodesics of test particles and photons are expected to be the same, providing a seemingly good BH mimicker. However, in this paper, we have shown that this is not always the case. 

In Fig.~\ref{fig:fig1} we have shown that rotating BS can develop the ergosphere outside its surface only if its dimensionless spin is rather high, i.e. $\beta > \beta_{*} \equiv 1/\sqrt{2} \sim 0.71$. This implies that the rotational energy of BS can be extracted only if the spin is above the critical $\beta_*$, i.e. the BS with the spin $\beta \leq \beta_*$  would be inefficient for any energy extraction mechanism, preventing the operation of many processes of particle acceleration and jet launching, including, e.g. \cite{Blandford1977, Tursunov20ApJ}. On the other hand, for spins $\beta_* < \beta \leq 1$ both BS and BH would have ergospheres, allowing efficient energy extraction of rotational energy.

It is interesting that BS can hold more rotation, $1 < \beta \leq 9/8$ than BH. This is perhaps because its surface is not null and hence may have some order of multipole moments, which may help in holding more spin. Hence the processes occurring at spins over-extremal with respect to BH are also expected to differ from that of the BH case. Moreover, a collapse of a BS with spin $\beta > 1$ can potentially produce an extremal Kerr BH with $\beta=1$ as neutral matter accretes. This may perhaps be the only way to form an extremal rotating BH as it has been shown that a non-extremal BH cannot be extremalized by adiabatic accretion \cite{Shaymatov23PLB}.  

Further, we investigated two energy extraction mechanisms from rotating BS: geometric Penrose process and magnetic Penrose process (MPP). In the former case we have found an analogue of the Wald inequality for the threshold relative velocity between the two fragments of the particle decay inside the ergosphere. While in the case of BH, the Wald inequality gives $v\geq 1/2$, for BS the condition is, as expected, even stronger, giving $v\geq \sqrt{0.3} \approx 0.547$. Therefore, the geometric Penrose process is not viable astrophysically for both BH and BS. However, all this gets circumvented when an electromagnetic interaction is involved, motivating us to study the MPP for BS.

MPP is fueled by the property that twisting of magnetic field lines around BH by the well-known frame dragging effect generates a quadruple electric field. 
\textcolor{black}{This electric field can be associated with the electric charge, which, in the case of the uniform magnetic field configuration, is known as the Wald charge and equals to $Q_{\rm W} = 2 \beta B$ \cite{Wald74}. Induced charge provides energy to one of the charged fragments to ride on negative energy orbit without making any demand on the relative velocity.} This is how energy extraction efficiency is enormously enhanced, making MPP the most attractive mechanism. 

\textcolor{black}{
It is pertinent to estimate the gravitational effect of the induced charge on the spacetime metric to verify the applicability of the Kerr solution, rather than the Kerr-Newman metric. A strong electric charge $Q_{\rm KN}$ that curves the spacetime metric around a rotating object must be of the order of its mass in geometrized units. In physical units (Gaussian and CGS-ESU), this gives $Q_{\rm KN} \approx 2 \sqrt{G} M$. In fact, it is exactly $2 \sqrt{G} M$ for a maximally charged non-rotating BH. Since BS can accumulate more charge than a BH, we use an approximate equality, which is sufficient for our estimation purposes. 
On the other hand, the Wald charge, after restoring the constants and expressing it in physical units, is of the order $Q_{\rm W} = 2 \beta B M^2 G^2/c^4$. Dividing one by another, we get 
\begin{equation}
    \frac{Q_{\rm W}}{Q_{\rm KN}} \approx \frac{G^{3/2}}{c^4} \beta B M \approx 4 \times 10^{-20} \,\,\, \frac{\beta}{1} \,\,\, \frac{M}{M_{\odot}} \,\,\, \frac{B}{1 {\rm Gauss}},  
\end{equation}
which implies that the gravitational effect of the induced charge is negligible in all astrophysically relevant scenarios, and the Kerr metric assumption is valid all through the energy extraction process, as the induced charge gets neutralized by the MPP. 
}

As the rotational energy of BH could be efficiently tapped by MPP, we have shown that the same is true for tapping the rotational energy of BS if the spin of the BS is above the critical value, i.e. $\beta > \beta_*$. As for the magnetic field source, we have studied two scenarios: uniform and dipole field. As shown in Fig.~\ref{fig:mpp} for the uniform field, the process is more efficient for BH than BS while for the dipole field, which is only applicable for BS, it is significantly more efficient than the uniform field case. In the former, the magnetic part of efficiency goes to zero for BH as $\beta\to1$ in contrast for BS it remains above $40\%$ for the extremal value $\beta=9/8$. For the dipole case, it always increases with increasing $\beta$.

Since BS is a horizonless compact object, the topology of the magnetic field on its surface may significantly differ from that of the event horizon scale of a corresponding BH. \textcolor{black}{While a BH cannot generate its own magnetic field, except for a magnetic monopole, a BS may potentially support electric currents on its surface. Therefore, in addition to the external magnetic field from surrounding plasma, a BS can develop its own dipolar or multipolar magnetic field, which may be considerably stronger than the typical fields expected from BH accretion processes.} Thus, a rapidly rotating BS with $\beta > \beta_*$ with its own strong magnetic field can be energetically much more efficient than the same mass BH. \textcolor{black}{Mapping the magnetic field structure at the horizon scale through Event Horizon Telescope (EHT) observations \cite{2024ApJ...964L..25E} offers a promising approach to distinguishing a BS from BH.}

Besides this, another distinguishing feature of BS is that there would always remain residual inextricable rotational energy corresponding to $\beta=\beta_*$. This is because rotational energy could only be extracted when there exists the ergoregion, which does not exist for $\beta\leq\beta_*$. The total rotational energy of an extremely rotating BH is $29\%$, which is fully extractable as the ergoregion exists for the entire range, $0<\beta\leq1$. This is therefore the maximum rotational energy available for extraction. On the other hand, from the rotating BS total extractable energy is $25\%$ corresponding to $\beta_*<\beta=9/8$. That is, rotational energy corresponding to $\beta\leq\beta_*$, which is $29-25 = 4\%$, remains as residue and cannot be extracted out.

\textcolor{black}{In order to distinguish BS from BH in observational data, it is necessary to perform measurements at spatial scales comparable to the gravitational radius of the object. This is becoming increasingly feasible due to advancements in observational techniques, particularly with the capabilities of the Very-Long-Baseline Interferometry (VLBI), including the EHT, and gravitational wave detectors, such as the Laser Interferometer Space Antenna (LISA). Extension of the VLBI to space will enable the direct measurements of the spin of candidate sources \cite{2024arXiv240612917J}. Spin measurements will also be achievable with the highly promising LISA mission, offering the potential for very high precision \cite{2004PhRvD..69h2005B}, including for near-extremal spin values \cite{2020PhRvD.102l4054B}. }

\textcolor{black}{BS with a spin below critical, $\beta\leq\beta_*$, will be energetically silent. The presence of a rotating compact object lacking signatures of energy extraction, such as powerful jets and energetic particle emissions, would provide a clear indication of the existence of a BS. }

\textcolor{black}{On the other hand, for higher spins, $\beta > \beta_*$, a BH and a BS would be energetically indistinguishable. However, since the surface of a BS can potentially emit electromagnetic radiation, like other stars, resolving and imaging the inner region of the object could offer additional insights. 
}

However, there is one serious caveat about using the Kerr metric for BS because it could truly describe only a rotating BH and not a non-BH rotating object. There exists no exact solution describing a rotating object without a horizon. This is in contrast with the static case where the metric describes the field of a static object irrespective of whether it is BH or not. Our justification for employing the Kerr metric for BS is that it is, though, not a BH but it is quite close to it as indicated by the defining condition, 
potential $\Phi(r) = 4/9$ for BS while it is $1/2$ for BH. Moreover, since a neutron star with the maximum stable mass on the verge of gravitational collapse to BH can be described in a good approximation by the Kerr metric \cite{2013MNRAS.433.1903U,2015PhRvD..92b3007C}, so can be the case for BS. The Kerr metric could therefore be taken as a good reasonable approximation. 

There is however one distinct advantage of having an object almost as compact as BH yet having a non-null timelike boundary which does not entirely block its interior from external interaction. There have been attempts to have a timelike fiducial surface very close to the horizon in membrane paradigm and stretched horizon proposal \cite{Susskind93PRD,Grumiller18IJMPD}. In BS we have a real astrophysical object that is almost as compact as BH having a timelike boundary. On this count too, BS is a welcome introduction in the study of physical phenomena in the BH vicinity.

\textcolor{black}{Future precise measurements of BH spins, combined with polarimetric data, will provide valuable insights that could further help in distinguishing between BH and BS, potentially offering a clearer understanding of their fundamental differences and the nature of compact objects.}

\section*{Acknowledgments}
This work is supported by the National Natural Science Foundation of China under Grants No. 11675143, the National Key Research and Development Program of China under Grant No. 2020YFC2201503, and the Czech Science Foundation (GA{\v C}R) Grant No.~23-07043S. N.D. wishes to acknowledge with thanks the visit to the University of KwaZulu-Natal, 
Durban. A.T. acknowledges the IUCAA for kind hospitality and the Alexander von Humboldt Foundation for the Fellowship. 

\appendix

\bibliographystyle{apsrev4-1}  
\bibliography{Ref_BS}

\end{document}